\documentclass[journal,10pt]{IEEEtran}
\usepackage{times,epsfig,epstopdf,latexsym,multirow,array,amssymb,graphicx,multicol,stfloats,textcomp,psfrag}
\usepackage[cmex10]{amsmath}
\usepackage{eqparbox,diagbox}
\usepackage{amsthm}
\usepackage{amsfonts}
\usepackage{here}
\usepackage{rawfonts}
\usepackage[latin1]{inputenc}
\usepackage[T1]{fontenc}
\usepackage{calc}
\usepackage{url}
\usepackage{enumerate}
\usepackage{color}
\usepackage{upref}
\usepackage{epic,eepic}
\usepackage{dsfont}
\usepackage{comment}
\usepackage{balance}
\usepackage{cite}
\usepackage{graphicx}

\theoremstyle{definition}

\usepackage{float}
\usepackage{stfloats}
\newcounter{tempequationcounter}
\usepackage{pdfpages}

\begin{document}

\title{EXIT Chart Analysis of Turbo Compressed Sensing Using Message Passing De-Quantization}

\author{Amin Movahed,~\IEEEmembership{Student Member,~IEEE,}
		Mark C. Reed,~\IEEEmembership{Senior Member,~IEEE,
		Shahriar Etemadi Tajbakhsh}
    \thanks{ 
%
   Amin Movahed, Mark C. Reed and Shahriar Etemadi Tajbakhsh are with the School of Engineering and Information Technology, University of New South Wales, Canberra, Australia (e-mails: \{a.movahed@student., mark.reed@, s.etemadiajbakhsh@\}unsw.edu.au\}).
}
\vspace{-9mm}
}

\maketitle

\IEEEpeerreviewmaketitle
\begin{abstract}
We propose an iterative decoding method, which we call turbo-CS, for the reception of concatenated source-channel encoded sparse signals transmitted over an AWGN channel. The turbo-CS encoder applies 1-bit compressed sensing  as a source encoder concatenated serially with a convolutional channel encoder. At the turbo-CS decoder, an iterative joint source-channel decoding method is proposed for signal reconstruction.
We analyze, for the first time, the convergence of turbo-CS decoder by determining an EXIT chart of the constituent decoders. We modify the soft-outputs of the decoder to improve the signal reconstruction performance of turbo-CS decoder. For a fixed signal reconstruction performance RSNR of $ \mathbf{10} $ dB we achieve more than $ \mathbf{5} $ dB of improvement in the channel SNR after $ \mathbf{6} $ iterations of the turbo-CS. Alternatively, for a fixed SNR of $ \mathbf{-1} $ dB, we achieve a $ \mathbf{10} $ dB improvement in RSNR.
\\
\\\emph{Index Terms}--- turbo-coding, 1-bit compressed sensing, joint source-channel decoding, message passing de-quantization
\end{abstract}
\section{Introduction}
Sparse signals can be widely used and observed in different applications from astronomical and medical imaging to data collected by a group of sensors \cite{bobin2008compressed,lustig2007sparse,haupt2008compressed}.
The sparsity in such signals is effectively exploited in compressed sensing (CS) to retrieve a high dimensional sparse signal via a smaller number of
measurements and to encode it to an arguably smaller number of values \cite{donoho2006compressed,candes2008introduction}. In other words, compressed sensing can be categorized as a source coding method 
for sparse signals. In many applications those sparse signals are required to be transmitted to a receiver over a communication channel which is typically prone 
to errors. Therefore channel coding is also needed to protect the transmission of the bits against the channel errors. It has been proven that under idealistic asymptotic conditions, where both source and channel codes are assumed to be of infinite length and complexity, optimal performance is achievable via separate source and channel coding. In practice, however, source coding usually leaves some redundancy in the representation of the signal. Joint source-channel decoding is provably optimum under such circumstances, where the residual redundancies from the source coding can be jointly used with redundancies introduced by the channel code to correct the channel-errors \cite{sayood1991use,guillemot2005joint}.

In this work, we consider a single-transmitter single-receiver communication system for transmission of sparse signals over an AWGN channel. We propose an iterative (turbo) method which we refer to as turbo-CS. The turbo-CS encoder consists of a 1-bit CS encoder, \cite{boufounos20081,jacques2011robust}, as a source encoder concatenated serially with a convolutional channel encoder. In other words, the \textit{inner} encoder in serial concatenated convolutional code is replaced with a 1-bit CS encoder. We introduce an iterative source-channel decoder where the process of decompression is performed within the loop of a turbo-decoder cycle. It should be noted that apart from the high performance of the proposed decoder due to the aforementioned reasons, the computational complexity vs. performance of our system is expected to be significantly better than a separate system of concatenated CS decoder and a turbo decoder.
 
 The term `turbo' refers to the approach of turbo-codes where two decoders at the receiver exchange \textit{a posteriori / a priori} information through a number of iterations to improve the decoding process \cite{berrou1993near,berrou1996near,hagenauer1995source}. Turbo-CS decoder applies an \textit{a posteriori probability} (APP) decoder as channel decoder and \textit{soft-in/soft-out} (SISO) 1-bit CS decoder as the source decoder.
 
 This paper is a substantial extension of our previous work in \cite{movahed2014iterative} where we have introduced the basic model of turbo-CS decoding system. In SISO 1-bit CS decoder in \cite{movahed2014iterative}, a heuristically chosen linear mapping function provides a priori information for APP decoder to be used in the next iteration of the turbo-CS decoder \cite{movahed2014iterative}. Since the proposed mapping method is based on a heuristic conversion function, its output bit probabilities are not of the proper distribution to be fed to the APP decoder as input. Consequently, despite the relatively good performance of the proposed system it is yet considerably sub-optimal. Also, \cite{movahed2014iterative} lacks a solid analysis on the performance of the system, again because of the heuristic mapping function used. 

The key contribution of this work is to introduce a new SISO 1-bit decoder which is based on message passing de-quantization (MPDQ) method in \cite{kamilov2012message}. MPDQ applies Gaussian approximation to solve a loopy belief propagation problem. The input of MPDQ is quantized version of noisy CS measurements. This method reconstructs the sparse signal by estimating the mean and variance of the signal elements from which the bit probabilities we are looking for can be derived. In a special case when the quantizer has two levels, the MPDQ algorithm converts to 1-bit CS reconstruction method using Gaussian approximation. We propose SISO 1-bit MPDQ algorithm whose input and output are soft-values. The SISO 1-bit MPDQ is a tractable algorithm and its output bit probabilities are matched with the required  bit probabilities by APP decoder. Therefore, we can analyse the performance and convergence of the turbo-CS decoder using extrinsic transfer (EXIT) chart. Moreover, we consider a more general class of Bernoulli-Gaussian signals in this work (in comparison to the model in \cite{movahed2014iterative}) where the sparsity level of the signal is also stochastic and unknown to the decoder.

This paper is organized as follows: a summary of related background is provided in Section \ref{rew}. In Section \ref{app}, we explain turbo-CS and channel transmission model. In Section \ref{deco}, we introduce turbo-CS decoder for the system model of Section \ref{app}. The EXIT chart analysis of the turbo-CS decoder is presented in Section \ref{ana}. The performance of turbo-CS through numerical experiment is investigated in Section \ref{per}. The paper is concluded in Section \ref{sec6}.
%
%
%
%

%

%
%
\section{Related works} \label{rew}
In this work, compressive sensing is applied in the context of joint source-channel decoding. CS has been introduced in \cite{donoho2006compressed,candes2005decoding} and extensively studied \cite{zhang2008compressed}. The number of CS measurements can be reduced in the presence of a statistical characterization of the signal where Bayesian inference offers the potential for more precise estimation of signal \cite{ji2008bayesian,seeger2008compressed}. In \cite{baron2010bayesian}, a belief propagation (BP) decoder has been developed to accelerate CS encoding and decoding under the Bayesian framework. To overcome the computational complexity of BP a Gaussian approximation is applied to CS decoding in \cite{rangan2010generalized}.   
 
In practical applications, the CS measurements are quantized to finite symbols for storage and transmission purposes \cite{laska2012regime,laska2011trust}. Gaussian approximation of BP, referred to MPDQ, has also been effectively applied in quantized compressed sensing \cite{kamilov2012message}. The SISO source decoding method introduced in our paper is inspired by MPDQ. However, we have modified MPDQ to accept and provide soft-values (probabilities) as input and output.

A wide range of joint source-channel decoding schemes has been proposed in the literature. Here, we list few mostly relevant works to ours (see \cite{guillemot2005joint} for an extended survey).
In \cite{schmalen2011exit}, EXIT chart analysis has been used to design an iterative source-channel decoder. 
Statistical characteristics of hidden Markov sources have been used to modify a turbo decoder for joint source-channel decoding \cite{garcia2001joint}.
Optimal vector quantizer has been incorporated as a channel encoder jointly with CS encoder for sparse signal transmission over AWGN channel \cite{shirazinia2014joint}. In a paper by Schniter, BP has been applied in an iterative approach for reconstruction of structured-sparse signals  \cite{schniter2010turbo}.

\section{System model} \label{app}
In this section we explain turbo-CS encoding model. We apply concatenated coding scheme consisting of the combination of two constituent encoders. Turbo-CS encoder includes a 1-bit CS encoder (as a source encoder) and a channel encoder to form a serial concatenated source-channel encoder. The input of turbo-CS encoder is a sparse signal.

\subsection{Sparse signal model}
We denote a block of sparse signal with $ \mathbf{x}\in \mathbb{R}^{N} $. Signal $ \mathbf{x} $ is called $ K $-sparse when $ K $ elements out of $ N $ elements are non-zero. In this work, we consider a case where $ K $ is stochastic and the probability of an element being non-zero is known and fixed to $ \rho $. We assume that the elements of $ \mathbf{x} $ are independent and identically distributed (i.i.d) and generated from the following Bernoulli-Gaussian distribution:
\begin{equation}
x_i\sim\begin{cases}
\mathcal{N}\left(0,\frac{1}{\rho}\right), & \textrm{with probability }\rho,\\
0, & \textrm{with probability }1-\rho
\end{cases}\label{dis}
\end{equation}
where $x_{i} $ denotes $ i $th element of vector $ \mathbf{x} $.
Therefore, the probability density function (PDF) of $ x_i $ is
\begin{equation}
\mathbb{P}_{x_{i}}\left(x\right)=\rho\sqrt{\frac{\rho}{2\pi}}e^{-\frac{\rho x^{2}}{2}}+\left(1-\rho\right)\delta\left(x\right). \label{dis2}
\end{equation}
The PDF of a signal block is obtained from
\begin{equation}
\mathbb{P}_{\mathbf{x}}\left(\mathbf{x}\right)=\prod_{i=1}^{N}\mathbb{P}_{x_{i}}\left({x}_{i}\right).
\end{equation}
  
\subsection{1-bit compressed sensing}
In classic CS, the sparse signal $ \mathbf{x} $ is measured through a few number of linear measurements.  The measurement vector $ \mathbf{y} $ is obtained by multiplying signal vector $ \mathbf{x} $ with matrix $ \boldsymbol{\Phi}\in \mathbb{R}^{M\times N} $. The elements of $ \boldsymbol{\Phi} $ are i.i.d. zero mean Gaussian random variables with variance $ 1/M $ and the output of CS is 
\begin{equation}
\mathbf{y}=\boldsymbol{\Phi}\mathbf{x} .
\end{equation} 
It is shown that matrix $ \boldsymbol{\Phi} $ satisfies the \textit{restricted isometry property} (RIP) which guarantees signal reconstruction with high probability \cite{candes2008restricted}. In quantized CS, for further storage and transmission purposes, the measurements are quantized to a number of alphabets. In addition, the noise of the measurement/quantization is modelled by an additive Gaussian noise \cite{laska2012regime}. We denote the measurement/quantization noise by vector $ \mathbf{e} $ where $e_{i}\sim\mathcal{N}\left(0,\sigma_{e}^{2}\right)$ for all $ i $. We denote a scalar quantization function with $\mathcal{Q}_L:\mathbb{R}\rightarrow\mathcal{F}$ that maps real-valued CS measurement to discrete alphabet $ \mathcal{F} $ with $\left|\mathcal{F}\right|=L$.

1-bit CS is a special case of quantized CS where $ L=2 $ and $\mathcal{F}=\left\{ -1,+1\right\} $. In fact, 1-bit quantizer is a simple sign function. We denote the block of the 1-bit CS measurements by $ \mathbf{b}\in\left\{ -1,+1\right\} ^{M} $ and 1-bit quantizer with $ \mathcal{Q}_2 $, hence, we have
\begin{equation}
\mathcal{Q}_2\left(\mathbf{y}+\mathbf{e}\right)=\textrm{sign}\left(\mathbf{y}+\mathbf{e}\right)=\mathbf{b}. \label{m1}
\end{equation}
Since in CS $ M<N $, it can be generally considered as a compression method. In particular, we are interested in 1-bit CS as the outputs are in the form of bits and therefore can be interfaced to conventional channel encoders.
\subsection{Concatenated source-channel encoder}
\begin{figure}[t]

\psfrag{x}[][]{$\mathbf{x}$}
\psfrag{y}[][]{$\mathbf{y}$}
\psfrag{b}[][]{$\mathbf{b}$}
\psfrag{d}[][]{$\mathbf{d}$}
\psfrag{A}[][]{{\small CS encoder}}
\psfrag{z}[][]{$\mathbf{z}$}
\psfrag{CS encoder}[][]{{\small CS encoder}}
\psfrag{B}[][]{$\textrm{sign}\left(\cdot\right)$}
\psfrag{C}[][]{{\small convolutional encoder}}
\psfrag{H}[][]{{\small AWGN}}
\psfrag{E}[][]{{\small 1-bit CS encoder}}
\hspace{-5mm}
\includegraphics[scale=1]{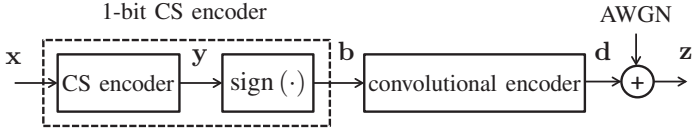}
\caption{Turbo-CS encoding: serial concatenation of 1-bit CS encoder and a convolutional encoder. Coded bits are transmitted over an AWGN channel\label{trans}}
\end{figure}

We consider the serial concatenated source/channel encoding transmission scheme of Figure \ref{trans}. The signal $ \mathbf{x} $ is compressed to $ \mathbf{b} $ by a 1-bit CS encoder with compression rate $ N/M $. We denote the bit energy by $ E_b $ as a measure of transmission power. In the next step, a channel code of rate $ M/P $ expands the block of 1-bit CS measurements $ \mathbf{b} $ to a sequence $ \mathbf{d} $ of size $ P $. 

In classic turbo encoder system models where two channel encoders are concatenated an interleaver is used between the two encoder at transmitter due to the correlation in the two encoding processes \cite{berrou1996near}. However, since in our scenario there is no correlation between source and channel coding, we confirmed through experiment that utilization of an interleaver does not improve the performance of the turbo system. In this work, we restrict our consideration to convolutional codes. The individual code bits $ y_i $ are BPSK-modulated with bit energy $ E_bM/P $ and transmitted over a memoryless AWGN channel with noise spectral density $ N_0=\sigma^2_c $.  

For simplicity, we assume that the measurement and quantization process in 1-bit CS is error-free and, therefore, there is no measurement/quantization noise in (\ref{m1}) ($ \sigma^2_e=0 $). As illustrated in Figure \ref{trans}, the only source of the noise in turbo-CS setup is the channel noise. 

The remainder of the paper is devoted to design and analysis of an iterative joint   source-channel decoder for the above transmission model. In the next section, we explain a turbo-decoder to reconstruct the signal from noisy turbo-CS encoded bits.

\section{Turbo-CS decoder} \label{deco} 
In this section, we propose an iterative joint source-channel decoder for mentioned transmission system in Section \ref{app} which we refer to as turbo-CS decoder. Turbo-CS decoder consists of two main parts: an APP decoder as the inner component decoder and a 1-bit CS decoder as the outer component decoder.  
\subsection{A posteriori probability decoder}
An APP decoder is a soft-in/soft-out (SISO) decoder \cite{benedetto1997soft}. The APP decoder accepts two inputs: 1) the received signal $ \mathbf{z} $ from the noisy channel and 2) the a priori probabilities of the bits. The latter is provided from the outer constituent decoder (in our case the 1-bit CS decoder) and in the form of soft-values. The soft-value is a convenient method of providing information about the sign and reliability of a bit. Specifically, the sign of a soft-value indicates the most likely sign of the bit and the magnitude gives the probability that it is correct. Soft-values can be in the form of log likelihood ratio (LLR) \cite{hagenauer1996iterative}. Given the probability of a bit, $ b\in\{-1,+1\} $, LLR of $ b $ is defined by
\begin{equation}
\Lambda:=\textrm{log}\left(\frac{\mathbb{P}_{b}\left(+1\right)}{1-\mathbb{P}_{b}\left(+1\right)}\right). \label{LLR}
\end{equation} 
The logarithm in (\ref{LLR}) is natural logarithm.
In the same way, the probability of $ b $ given $ \Lambda $ is calculated by
\begin{equation}
\mathbb{P}_{b}\left(+1\right)=\left(\frac{\textrm{tanh}\left(\frac{\Lambda}{2}\right)+1}{2}\right). \label{L2p}
\end{equation}

APP decoder applies BCJR  algorithm which generates APP of the states and transitions of a discrete time finite-state Markov process \cite{bahl1974optimal}. The output of APP decoder is in the form of extrinsic LLR values. We denote the vector of extrinsic LLR values at the output of APP decoder with $ \Lambda_{\textrm{CD}}^{[\textrm{ext}]}$. In addition, $ \Lambda_{\textrm{CD}}^{[\textrm{apri}]}$ refers to a priori LLR values at the input of APP decoder (channel decoder).  

The aim with the turbo-CS decoder is to apply a SISO inner component decoder that updates the extrinsic LLR values, $ \Lambda_{\textrm{CD}}^{[\textrm{ext}]}$, from APP decoder. Then the updated LLR values are passed to APP decoder as a priori LLR values, $ \Lambda_{\textrm{CD}}^{[\textrm{apri}]}$, for the next decoding iteration.

Thus, the inner decoding component should be in the form of SISO 1-bit CS decoder. In the following section, we propose a SISO 1- bit CS decoder as the inner decoder component of turbo-CS decoder.

\subsection{Bayesian formulation of soft-in/soft-out 1-bit CS decoding problem\label{BP}} 
In this section, we derive Bayesian formulation of SISO 1-bit CS decoding problem for turbo-CS decoding. In the first step, we focus on a transmission case where there is no noise in the channel. Therefore, the convolutional encoder in Figure \ref{trans} can be removed and we have $ \mathbf{z} =\mathbf{b}$ where $ \mathbf{b} $ is obtained from (\ref{m1}). As mentioned in the previous section, we assume that the measurement/quantization process is error-free and $ \sigma^2_e=0 $.  
Hence, conditional PDF of the signal $ \mathbf{x} $ given the received $ \mathbf{b} $ is 

\begin{eqnarray}
\mathbb{P}_{\mathbf{x}|\mathbf{b}}\left(\mathbf{x}|\mathbf{b}\right) & \varpropto & \mathbb{P}_{\mathbf{b}|\mathbf{y}}\left(\mathbf{b}|\mathbf{y}\right)\mathbb{P}_{\mathbf{x}}\left(\mathbf{x}\right)\\
 & \varpropto & \prod_{i=1}^{M}\mathbb{P}_{b|y}\left(b_{i}|y_{i}\right)\prod_{i=1}^{M}\mathbb{P}_{x_i}\left(x_{i}\right)
\label{dist}
\end{eqnarray}
where $ \varpropto $ denotes equality after normalization to unity. Since there is no measurement/quantization noise, we have

\begin{equation}
\mathbb{P}_{b|y}\left(b_{i}|y_{i}\right)=\begin{cases}
1, & \textrm{if}\, y_{i}\in Q_2^{-1}\left(b_{i}\right),\\
0, & y_{i}\notin Q_2^{-1}\left(b_{i}\right).
\end{cases}
\end{equation}
where 

\begin{eqnarray}
Q_2^{-1}\left(b\right) & = & \begin{cases}
\left[0,+\infty\right), & \textrm{if }b=+1,\\
\left(-\infty,0\right), & b=-1.
\end{cases}
\end{eqnarray}

The distribution in (\ref{dist}), describes the complete statistical characterization of the decoding problem. The minimum mean square error (MMSE) estimator of $ \mathbf{x} $ is obtained from the following

\begin{equation}
\hat{\mathbf{x}}_{\textrm{MMSE}}\left(\mathbf{b}\right)=\mathbb{E}\left(\mathbf{x}|\mathbf{b}\right).
\label{MMSE}
\end{equation}

As shown in Figure \ref{trans}, due to the channel noise there is uncertainty in received signal $ \mathbf{z} $ and after channel decoding, the APP decoder provides probability of bits instead of fixed $ \mathbf{b} $. We have 
\begin{equation}
\mathbb{P}^{[\textrm{ext}]}_{\mathbf{b}}\left(\mathbf{b}\right)=\prod_{i=1}^{M}\mathbb{P}^{[\textrm{ext}]}_{b_{i}}\left(b_{i}\right)
\label{bprob}
\end{equation}
where $ \mathbb{P}^{[\textrm{ext}]}_{b_{i}}(b_i)$ for each element of $ \mathbf{b} $ is obtained from (\ref{L2p}).

Therefore, the decoding problem for the SISO 1-bit CS decoder is the estimation of $ \mathbf{x} $ given the bit probabilities in (\ref{bprob}). From the law of total expectation we have
\begin{equation}
\mathbb{E}\left(\hat{\mathbf{x}}_{\textrm{MMSE}}(\mathbf{b})\right)=\sum_{\textrm{all }\mathbf{b}}\mathbb{E}\left(\mathbf{x}|\mathbf{b}\right)\mathbb{P}^{[\textrm{ext}]}_{\mathbf{b}}\left(\mathbf{b}\right)=\mathbb{E}\left(\mathbf{x}\right)=\hat{\mathbf{x}}
\label{MM}
\end{equation}
where $ \hat{\mathbf{x}} $ denotes the estimate of original signal $ \mathbf{x} $ via 1-bit SISO decoding. In other words, in turbo-CS decoder, the output LLR-values of APP decoder (\ref{LLR}) are input as soft-values to SISO 1-bit CS decoder. As the next step, SISO 1-bit CS decoder estimates original signal $ \mathbf{x} $ from (\ref{MM}) through an estimation method. A popular iterative computational method to approximate the MMSE estimator $ \hat{\mathbf{x}}_{\textrm{MMSE}} $ is loopy belief propagation (LBP) \cite{bishop2006pattern}. However, implementation of LBP for de-quantization problems when matrix $ \boldsymbol{\Phi} $ is not sparse is highly computationally complex. LBP requires computation of high-dimensional distributions at measurement nodes which makes its application impractical in de-quantization problems.
In the next section, we explain SISO 1-bit CS decoding algorithm which solves (\ref{MM}) using Gaussian approximate message passing technique \cite{rangan2010generalized,donoho2009message,boutros2002iterative}.
\subsection{Soft-in/soft-out 1-bit message passing de-quantization}
As discussed in section \ref{BP}, due to high complexity, implementation of MMSE estimator in (\ref{MMSE}) using belief propagation techniques is not possible. Recently, Kamilov \textit{et al.} introduced message passing de-quantization (MPDQ) algorithm to solve (\ref{MMSE}) \cite{kamilov2012message}. MPDQ is based on the Gaussian approximate message-passing algorithm referred to as generalized approximate message passing (G-AMP) \cite{rangan2010generalized}. 

In the following, we modify MPDQ to estimate the original signal $ \mathbf{x} $ to $ \hat{\mathbf{x}} $ in (\ref{MM}). The proposed SISO 1-bit CS decoder (as the inner component of turbo-CS decoder) applies the modified version of MPDQ to estimate the original signal $ \mathbf{x} $ jointly with APP decoder through turbo-iterations.

We have modified two parts in MPDQ:
\begin{itemize}
\item We change the non-linear factor update functions and adapt them to accept the probability of the bits as input. The input bit probabilities $ \mathbb{P}^{[\textrm{ext}]}_{\mathbf{b}}(\mathbf{b}) $ are obtained from a priori LLR values given to SISO 1-bit CS decoder denoted by vector  $ \boldsymbol{\Lambda}_{\textrm{SD}}^{[\textrm{apri}]}$ (a priori LLR values for source decoder).
\item We extend the algorithm to provide bit probabilities at the output of MPDQ (for the a priori input of the APP decoder on the next iteration). We denote the related LLR values at the output of MPDQ by vector $ \boldsymbol{\Lambda}_{\textrm{SD}}^{[\textrm{ext}]}$.
\end{itemize}
Here, we explain MPDQ algorithm including the modified steps:    
\subsubsection*{\textbf{1) Initialization}} The initial expected and variance vector are set with respect to prior distribution of $ \mathbf{x} $ in (\ref{dis2}). Therefore, $ \hat{\mathbf{x}}(0)=\mathbb{E}\left(\mathbf{x}\right)=\mathbf{0} $ and   $ \mathbf{v}^{x}(0)=\textrm{var}\left(\mathbf{x}\right)=\mathbf{1} $ where the number inside the parenthesis denotes the number of corresponding iteration and $ \mathbf{0} $ and $ \mathbf{1} $ denotes all-zero and all-one vectors respectively. We set initial value for the intermediate variable $ \hat{\mathbf{s}}(0)=\mathbf{0} $.
\subsubsection*{\textbf{2) Linear factor update functions}}

\begin{equation}
\mathbf{v}^{p}\left(t\right)=\left(\mathbf{\boldsymbol{\Phi}\bullet\boldsymbol{\Phi}}\right)\mathbf{v}^{x}\left(t-1\right) 
\end{equation}
\begin{equation}
\hat{\mathbf{p}}\left(t\right)=\boldsymbol{\Phi}\hat{\mathbf{x}}\left(t-1\right)-\mathbf{v}^{p}\left(t\right)\bullet\hat{\mathbf{s}}\left(t-1\right)
\end{equation}
\subsubsection*{\textbf{3) Non-linear factor update functions}}
\begin{equation}
\hat{s}_{i}\left(t\right)=\mathcal{E}_{\textrm{F}}\left(\mathbb{P}^{[\textrm{ext}]}_{b_{i}}\left(b_{i}\right),\hat{p}_{i}\left(t\right),v_{i}^{p}\left(t\right)\right)
\end{equation}
\begin{equation}
v_{i}^{s}\left(t\right)=\mathcal{V}_{\textrm{F}}\left(\mathbb{P}^{[\textrm{ext}]}_{b_{i}}\left(b_{i}\right),\hat{p}_{i}\left(t\right),v_{i}^{p}\left(t\right)\right)
\end{equation}
where $\bullet$ denotes element-wise multiplication and $ \mathcal{E}_{\textrm{F}} $ and $ \mathcal{V}_{\textrm{F}} $ are functions over scalar values:
\begin{equation}
\mathcal{E}_{\textrm{F}}\left(\mathbb{P}_{b}(b),\hat{p},v^{p}\right)=\frac{1}{v^{p}}\left(\mathbb{E}\left(z\right)-\hat{p}\right) \label{exp}
\end{equation}

\begin{equation}
\mathcal{V}_{\textrm{F}}\left(\mathbb{P}_{b}(b),\hat{p},v^{p}\right)=\frac{1}{v^{p}}\left(1-\frac{\textrm{var}\left(z\right)}{v^{p}}\right) \label{var}
\end{equation}
where we have a priori distribution $z\sim\mathcal{N}\left(\hat{p},v^{p}\right)$. The non-linear functions $ \mathcal{E}_{\textrm{F}} $ and $ \mathcal{V}_{\textrm{F}} $ are the modified versions of updating factor functions in \cite{kamilov2012message}. In fact, these modified functions allow a more general input model where all quantization symbols for each factor are possible as input with different probabilities. The expectation and variance in (\ref{exp}) and (\ref{var}) are calculated as follows: from the law of total expectation we have

\begin{equation}
\mathbb{E}\left(z\right)=\sum_{b=-1,+1}\mathbb{P}_{b}\left(b\right)\mathbb{E}\left(z|z\in Q_2^{-1}(b)\right).
\label{texp}
\end{equation}
In addition, from the law of total variance we obtain
\begin{eqnarray}
\textrm{var}\left(z\right) & = & \mathbb{E}_{b}\left(\textrm{var}\left(z|b\right)\right)+\textrm{var}_{b}\left(\mathbb{E}\left(z|b\right)\right)=\nonumber \\
 & = & \mathbb{E}_{b}\left(\textrm{var}\left(z|b\right)\right)+\mathbb{E}_{b}\left(\mathbb{E}\left(z|b\right)^{2}\right)-\mathbb{E}\left(z\right)^{2}
\label{tvar}
\end{eqnarray}
where

\begin{equation}
\mathbb{E}_{b}\left(\textrm{var}\left(z|b\right)\right)=\sum_{b=-1,+1}\mathbb{P}_{b}\left(b\right)\textrm{var}\left(z|z\in Q_2^{-1}(b)\right)
\label{texp2}
\end{equation}

\begin{equation}
\mathbb{E}_{b}\left(\mathbb{E}\left(z|b\right)^{2}\right)=\sum_{b=-1,+1}\mathbb{P}_{b}\left(b\right)\mathbb{E}\left(z|z\in Q_2^{-1}(b)\right)^{2}.
\label{texp3}
\end{equation}

We refer the reader to \cite{johnson1995continuous} for calculation of the expectation and variance of the truncated Gaussian distributions in (\ref{texp}), (\ref{texp2}) and (\ref{texp3}).

\begin{figure*}[!b]
 \vspace{-0.4cm}
 \normalsize
 \hrulefill
 \setcounter{tempequationcounter}{\value{equation}}
 \setcounter{equation}{33}
 \begin{equation}
\mathcal{I}\left(b;\Lambda\right)=\frac{1}{2}\sum_{b=-1,+1}\mathbb{P}_{\Lambda|b}\left(\Lambda|b\right)\log_{2}\left(\frac{2\mathbb{P}_{\Lambda|b}\left(\Lambda|b\right)}{\mathbb{P}_{\Lambda|b}\left(\Lambda|-1\right)+\mathbb{P}_{\Lambda|b}\left(\Lambda|+1\right)}\right)\label{MI}
\vspace{-10mm}
 \end{equation}
\setcounter{equation}{\value{tempequationcounter}}
 \vspace*{4pt}
 \end{figure*}
 
\subsubsection*{\textbf{4) Linear variable update functions}}

\begin{equation}
\mathbf{v}^{r}\left(t\right)=\left(\left(\boldsymbol{\Phi}\bullet\boldsymbol{\Phi}\right)^{T}\mathbf{v}^{s}\left(t\right)\right)^{-1}
\end{equation}

\begin{equation}
\hat{\mathbf{r}}\left(t\right)=\hat{\mathbf{x}}\left(t-1\right)+\mathbf{v}^{r}\left(t\right)\bullet\left(\boldsymbol{\Phi}^{T}\hat{\mathbf{s}}\left(t\right)\right)
\end{equation}
\subsubsection*{\textbf{5) Non-linear variable update functions}}
\begin{equation}
\hat{x}_{i}\left(t\right)=\mathcal{E}_{\textrm{V}}\left(\hat{r}_{i}\left(t\right),v_{i}^{r}\left(t\right)\right)
\end{equation}
\begin{equation}
v_{i}^{x}\left(t\right)=\mathcal{V}_{\textrm{V}}\left(\hat{r}_{i}\left(t\right),v_{i}^{r}\left(t\right)\right) 
\end{equation}
where  $ \mathcal{E}_{\textrm{V}} $ and $ \mathcal{V}_{\textrm{V}} $ are functions over scalar values and we have
\begin{equation}
\mathcal{E}_{\textrm{V}}\left(\hat{r},v^{r}\right)=\mathbb{E}\left(x|\hat{r}\right) 
\label{expi}
\end{equation}
and
\begin{equation}
\mathcal{V}_{\textrm{V}}\left(\hat{r},v^{r}\right)=\textrm{var}\left(x|\hat{r}\right).
\label{vari}
\end{equation}
The above expected value and variance are with respect to $ \hat{r}=x+w $ where $ w\sim \mathcal{N}(0,v^{r}) $. 

\subsubsection*{\textbf{6) Termination of iterations}}  Then, $t\leftarrow t+1$ and proceed to step 2) until convergence criterion is satisfied or $ t $ reaches the maximum iteration number. Therefore, through above updating steps MPDQ estimates $ \hat{\mathbf{x}} $ of original signal $ \mathbf{x} $ through inner iterations.
\subsubsection*{\textbf{7) Soft-output}} 
The term 'soft-output' refers to the updated bit probabilities at the output of SISO 1-bit CS decoder. To update the bit probabilities, we need to find the distribution of the estimate of $ \mathbf{y} $. We can approximate the distribution of the estimated elements of original signal $ \mathbf{x} $ with mean and variance of $ \mathbf{x} $ in (\ref{expi}) and (\ref{vari}). Since $ \mathbf{y} $ is a linear combination of $ \mathbf{x} $, the estimate of original $ \mathbf{y} $ is obtained from
\begin{equation}
\hat{\mathbf{y}}=\mathbb{E}\left(\mathbf{y}\right)=\boldsymbol{\Phi}\hat{\mathbf{x}}. \label{ymean}
\end{equation}
and
\begin{equation}
\mathbf{v}^{y}=\textrm{var}\left(\mathbf{y}\right)=\left(\boldsymbol{\Phi}\bullet\boldsymbol{\Phi}\right)\mathbf{v}^{x}
\label{yvar}
\end{equation}
where each element of the estimated $ \mathbf{y} $ has approximately Gaussian distribution with the mean and variance in \eqref{ymean} and \eqref{yvar}. Hence, we can update the probability of each corresponding bit at the output of SISO 1-bit CS decoder which is denoted by $ \mathbb{P}_{\mathbf{b}}^{[\textrm{apri}]} $.  The updated bit probabilities are calculated from

\begin{equation}
\mathbb{P}_{b_{i}}^{[\textrm{apri}]}\left(b_{i}\right)=\textrm{Q}\left(\frac{-b_{i}\hat{y}_{i}}{\sqrt{v_{i}^{y}}}\right). \label{qfun}
\end{equation}
where $ \textrm{Q}\left(x\right)=\frac{1}{\sqrt{2\pi}}\int_{x}^{\infty}\textrm{exp}\left(\frac{-u^{2}}{2}\right)du $.
Then the updated bit probabilities are converted to LLR values via \eqref{LLR} and form vector $ \boldsymbol{\Lambda}_{\textrm{SD}}^{[\textrm{ext}]}$ (extrinsic LLR values of source decoder).  We refer to the mentioned algorithm as SISO 1-bit MPDQ. 
\subsection{Combination of APP decoding and soft-in/soft-out 1-bit message passing de-quantization}
\begin{figure}[]
\centering
\psfrag{A}[][]{APP Decoder}
\psfrag{B}[][]{$ \mathbf{z} $}
\psfrag{P}[][]{Probability}
\psfrag{C}[][]{Transformer}
\psfrag{M}[][]{Modified}
\psfrag{F}[][]{Factor}
\psfrag{U}[][]{Update}
\psfrag{V}[][]{Variable}
\psfrag{T}[][]{soft-output}
\psfrag{W}[][]{generator}
\psfrag{E}[][]{$ \boldsymbol{\Lambda}_{\textrm{SD}}^{[\textrm{ext}]}$}
\psfrag{R}[][]{$ \boldsymbol{\Lambda}_{\textrm{SD}}^{[\textrm{mod}]}$}
\psfrag{L}[][]{$ \boldsymbol{\Lambda}_{\textrm{CD}}^{[\textrm{ext}]}$}
\psfrag{Z}[][]{$ \boldsymbol{\Lambda}_{\textrm{CD}}^{[\textrm{apri}]}$}
\psfrag{S}[][]{$ \boldsymbol{\Lambda}_{\textrm{SD}}^{[\textrm{apri}]}$}
\psfrag{X}[][]{$ \hat{\mathbf{x}} $}
\psfrag{G}[][]{SISO 1-bit MPDQ}

\includegraphics[scale=1]{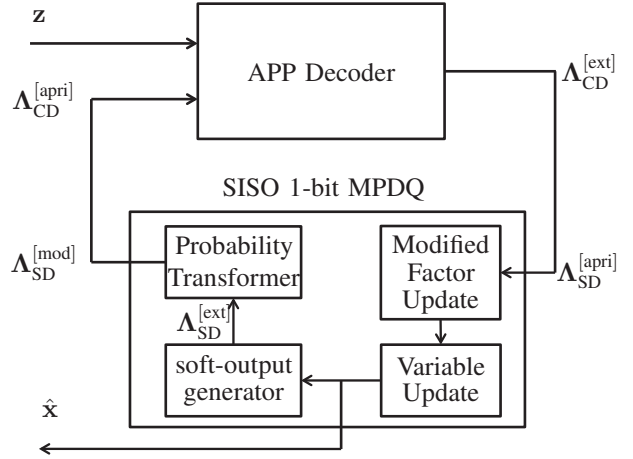}\caption{Turbo-CS decoding: APP decoder and SISO 1-bit MPDQ decoder are iteratively estimate signal $ \hat{\mathbf{x}} $ by exchanging LLR values \label{dia}}
\end{figure}

In Turbo-CS decoding, APP decoder and  SISO 1-bit MPDQ are applied as inner and outer decoder components respectively. In each iteration of turbo-CS decoder, the APP decoder decodes each block of received signal $ \mathbf{z} $ to provide extrinsic LLR values, $ \boldsymbol{\Lambda}_{\textrm{CD}}^{[\textrm{ext}]}$. These extrinsic values are given to SISO 1-bit MPDQ as $ \boldsymbol{\Lambda}_{\textrm{SD}}^{[\textrm{apri}]}$. SISO 1-bit MPDQ through number of inner iterations estimate signal block $ \hat{\mathbf{x}} $ and updates bit probabilities from \eqref{qfun}. The updated bit probabilities $ \mathbb{P}_{\mathbf{b}}^{[\textrm{apri}]}\left(\mathbf{b}\right) $ are converted to $ \boldsymbol{\Lambda}_{\textrm{SD}}^{[\textrm{ext}]}$ and given to APP decoder as $ \boldsymbol{\Lambda}_{\textrm{CD}}^{[\textrm{apri}]}$ for the next turbo-iteration.
Turbo-CS decoder, estimates original signal $ \mathbf{x} $ with $ \hat{\mathbf{x}} $ through several turbo-iterations (Figure \ref{dia}). In the first iteration of turbo-CS, there is no a priori information available at the input of APP decoder and  a priori bits are equiprobable, hence, $ \boldsymbol{\Lambda}_{\textrm{CD}}^{[\textrm{apri}]}$ is initialized with all-zero vector.

\section{Analysis of turbo-CS decoder}\label{ana}
\subsection{EXIT chart for iterative processes} \label{A}
Extrinsic information transfer (EXIT) chart is a popular and powerful tool for analysis and prediction of the convergence behaviour of iterative techniques \cite{ten2001convergence,ten1999convergence,hagenauer2004exit}. In this section, we briefly review the EXIT chart approach. We refer the reader to \cite{ten2001convergence} for comprehensive explanation of EXIT chart.

In an EXIT chart, each constituent decoder of the turbo process is presented by an EXIT function. The EXIT function describes the relation between input/output of each individual decoder by applying mutual information measure $ \mathcal{I}\left(\cdot;\cdot\right) $ from \eqref{MI}.   
For constituent decoders of turbo-CS scheme we have:
\begin{itemize}
\item APP decoder:
\[
\mathcal{I}_{\textrm{CD}}^{[\textrm{apri}]}=\mathcal{I}\left(b;\Lambda_{\textrm{SD}}^{[\textrm{ext}]}\right)\,\rightarrow\,\mathcal{I}_{\textrm{CD}}^{[\textrm{ext}]}=\mathcal{I}\left(b;\Lambda_{\textrm{CD}}^{[\textrm{ext}]}\right)
\]

\item SISO 1-bit MPDQ decoder:
\[
\mathcal{I}_{\textrm{SD}}^{[\textrm{apri}]}=\mathcal{I}\left(b;\Lambda_{\textrm{CD}}^{[\textrm{ext}]}\right)\,\rightarrow\,\mathcal{I}_{\textrm{SD}}^{[\textrm{ext}]}=\mathcal{I}\left(b;\Lambda_{\textrm{SD}}^{[\textrm{ext}]}\right)
\]

\end{itemize}

\begin{figure}[]
\psfrag{A}[][]{$ \mathcal{I}_{\textrm{CD}}^{[\textrm{apri}]} $, $ \mathcal{I}_{\textrm{SD}}^{[\textrm{ext}]} $ [bit]}
\psfrag{B}[][]{$ \mathcal{I}_{\textrm{SD}}^{[\textrm{apri}]} $, $ \mathcal{I}_{\textrm{CD}}^{[\textrm{ext}]} $ [bit]}
\hspace{-0.7cm}
\includegraphics[scale=1]{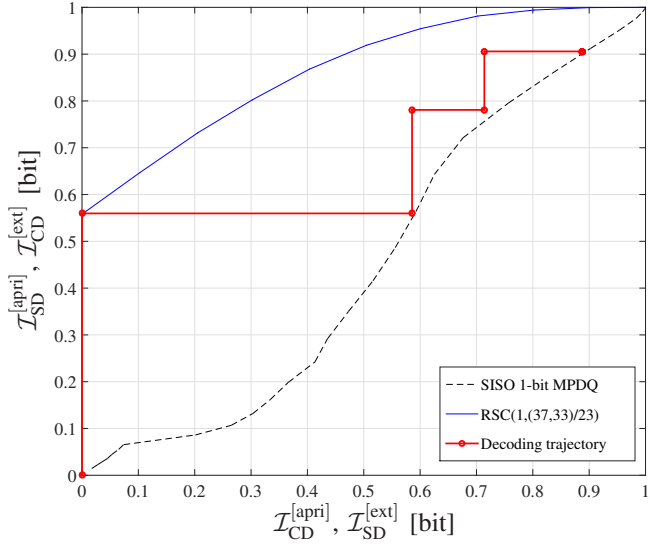}\caption{EXIT chart of turbo-CS receiver at SNR$=-1 $ [dB] \label{EX1}}
\end{figure}

The EXIT functions for 1-bit MPDQ and APP decoder are illustrated in Figure \ref{EX1}. Here, we apply a corresponding APP decoder of a rate $ 1/3 $ recursive systematic convolutional channel encoder . The code polynomials are $\mathbf{G}=\left(1,\frac{37}{23},\frac{33}{23}\right)_{8}$. Note that the EXIT function for APP decoder depends on the quality of channel because $ \mathbb{P}_{z|d}(z|d) $ is considered in determination of $ \Lambda_{\textrm{CD}}^{[\textrm{ext}]} $ \cite{hagenauer2004exit}. We define signal to noise ratio (SNR) as the channel noise measure by
\begin{equation}
\textrm{SNR}=\frac{\textrm{E}_{b}}{\textrm{N}_{0}}=\frac{1}{\sigma_c^{2}}
\end{equation}
where $ \textrm{E}_{b} $ is the average power of a bit at the output of 1-bit CS encoder.  We set $\textrm{SNR}=-1 $ dB and bit block size $ M=500 $. The EXIT functions are measured individually while the decoding trajectory is measured in the actual turbo-CS configuration. In Figure \ref{EX1}, the decoding trajectory is shown for $ 3 $ turbo-iterations by the circled line. The decoding tunnel between two EXIT functions indicates that turbo-CS decoder might converge after number of iterations. 

As it can be seen, the decoding trajectory undershoots the APP EXIT curve. In classic EXIT chart, the undershooting (overshooting) effect can result from a too small signal block size \cite{ten2001convergence}. However, this is not the case in turbo-CS scheme as there is no correlation between the inner and outer component coding systems. Hence, there is no interleaver applied in turbo-CS and the signal block length does not affect the turbo-decoding performance. In our case, the undershooting effect is due to the mismatch between the output LLR values of SISO 1-bit MPDQ, $ \Lambda_{\textrm{SD}}^{[\textrm{ext}]} $, and input LLR values of APP decoder, $ \Lambda_{\textrm{CD}}^{[\textrm{apri}]} $. In fact, the EXIT functions in Figure \ref{EX1} are obtained though individual classic configuration where $ \Lambda_{\textrm{CD}}^{[\textrm{apri}]} $ and $ \Lambda_{\textrm{SD}}^{[\textrm{apri}]} $ are modelled with the following random process  :
\begin{equation}
\setcounter{tempequationcounter}{\value{equation}}
 \setcounter{equation}{35}
\Lambda=\mu_{\Lambda}b+n_{\Lambda} \label{mo}
\end{equation}
where $ n_{\Lambda}\sim\mathcal{N}\left(0,\sigma_{\Lambda}^{2}\right) $ and $ \mu_\Lambda=\sigma_{\Lambda}^{2}/2 $ \cite{ten2001convergence}.
However, the obtained $ \Lambda_{\textrm{SD}}^{[\textrm{ext}]} $ in actual configuration (trajectory) does not have the same random Gaussian characteristic as in \eqref{mo}. The reason behind this mismatch is the approximation applied in  generation of soft-output values in SISO 1-bit MPDQ in \eqref{ymean}, \eqref{yvar} and \eqref{qfun}.

In the next section, a training-based modification method will be introduced which maps $ \Lambda_{\textrm{SD}}^{[\textrm{ext}]} $ to true LLR values.

\subsection{Modification of a priori LLR values} \label{e}
To obtain true LLR values from $ \Lambda_{\textrm{SD}}^{[\textrm{ext}]} $, we consider the following characteristic of LLR values generated from \eqref{mo} 
\begin{equation}
\Lambda=\textrm{log}\left(\frac{\mathbb{P}_{\Lambda|b}\left(\Lambda|+1\right)}{\mathbb{P}_{\Lambda|b}\left(\Lambda|-1\right)}\right). \label{Mod}
\end{equation}

This property of LLR values gives us a benchmark to investigate the accuracy of the obtained probabilities from (\ref{qfun}). In other words, if the corresponding bit probabilities, $ \mathbb{P}_{b}^{[\textrm{apri}]}(b), $ were true then $ \Lambda_{\textrm{SD}}^{[\textrm{ext}]} $ would satisfy the property in (\ref{Mod}). 
\begin{figure}[t]
\hspace{-.5cm}
\psfrag{A}[][]{$ \mathbb{P}_{b}^{[\textrm{apri}]}\left(b\right) $}
\psfrag{B}[][]{$ \mathbb{P}_{b}^{[\textrm{mod}]}\left(b\right) $}
\psfrag{Practical}[][]{{\tiny Genie aided}}
\psfrag{Tanh}[][r]{{\tiny Hyperbolic}}
\includegraphics[scale=1]{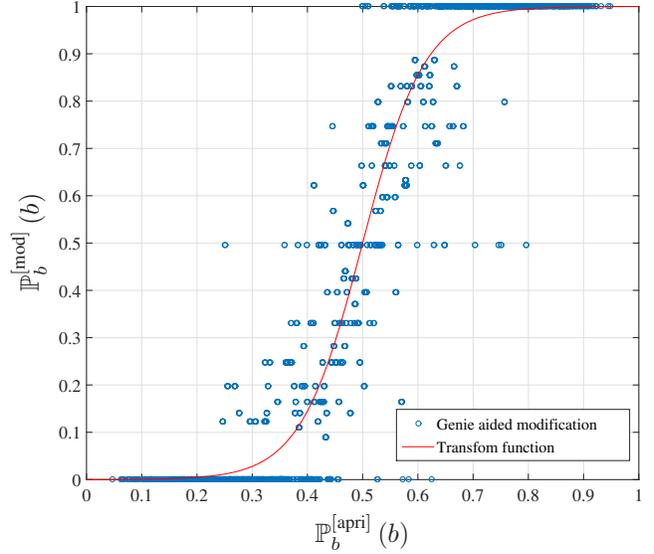}\caption{Input probability vs. modified probability at the output comparing the Genie aided results with our proposed transform function (M=500, T=5) \label{hy}}
\end{figure}

\begin{figure}[t]
\hspace{-.5cm}
\psfrag{A}[][]{$ \mathbb{P}_{b}^{[\textrm{apri}]}\left(b\right) $}
\psfrag{B}[][]{$ \mathbb{P}_{b}^{[\textrm{mod}]}\left(b\right) $}
\psfrag{Practical}[][]{{\tiny Genie aided}}
\psfrag{Tanh}[][r]{{\tiny Hyperbolic}}
\includegraphics[scale=1]{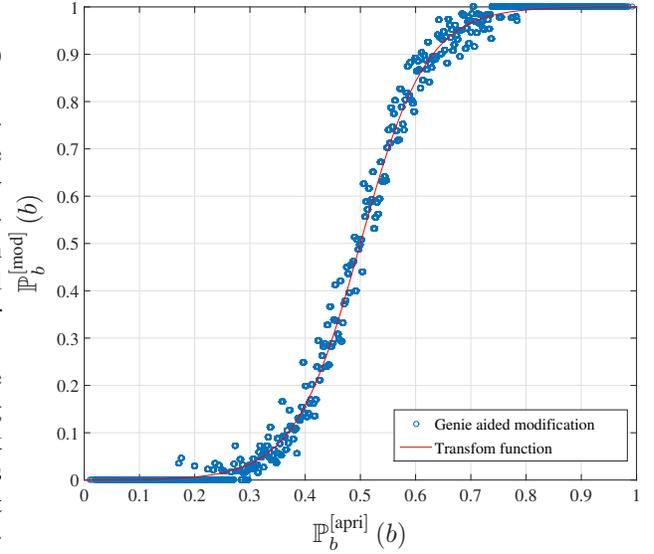}\caption{Input probability vs. modified probability at the output comparing the Genie aided results with our proposed transform function (M=500, T=50)\label{hy2}}
\end{figure}

Accordingly, we calculate the right-hand side of (\ref{qfun}) and denote the result LLR values (left-hand side) by $ \Lambda_{\textrm{SD}}^{[\textrm{mod}]} $. The obtained values of $ \Lambda_{\textrm{SD}}^{[\textrm{mod}]} $ are true LLR values that need to be given to APP decoder as $ \Lambda_{\textrm{CD}}^{[\textrm{apri}]} $. Determining the conditional probability values in \eqref{Mod} requires knowledge of original output bits of 1-bit CS encoder, $ \mathbf{b} $, at receiver. Here, we apply a $ T $ block of training bit sequences which is predefined to both transmitter and receiver parts. For simplicity, we derive a modification function for bit probabilities rather than LLR values as probabilities have a bounded magnitude. The flow of modification process is as follows:
\[
\mathbb{P}_{b}^{[\textrm{apri}]}\left(b\right)\overset{(1)}{\longrightarrow}\Lambda_{\textrm{SD}}^{[\textrm{ext}]}\overset{(2)}{\longrightarrow}\Lambda_{\textrm{SD}}^{[\textrm{mod}]}\overset{(3)}{\longrightarrow}\mathbb{P}_{b}^{[\textrm{mod}]}\left(b\right)
\]
\begin{itemize}
\item[(1)] The obtained bit probabilities in \eqref{qfun} are mapped to $ \Lambda_{\textrm{SD}}^{[\textrm{ext}]} $ from \eqref{LLR}. 
\item[(2)] $ \Lambda_{\textrm{SD}}^{[\textrm{ext}]} $ are modified to $ \Lambda_{\textrm{SD}}^{[\textrm{mod}]} $ by using training bit sequence in \eqref{Mod}. The conditional probabilities in \eqref{Mod} are calculated from samples $ \Lambda_{\textrm{SD}}^{[\textrm{ext}]} $ and training bits. In this fashion, $ \Lambda_{\textrm{SD}}^{[\textrm{ext}]} $ are grouped in a number of histogram bins. Therefore, the precision of conditional probabilities depends on the number of training blocks, $ T $ (Figure \ref{hy2} and \ref{hy}).
\item [(3)] Modified LLRs are mapped to modified bit probabilities from \eqref{L2p}.
\end{itemize}

The modification scatter diagram of bit probabilities is illustrated in Figure \ref{hy2} and \ref{hy} for the first iteration of turbo-CS setup in Section \ref{A} and for $ T=5 $ and $ T=50 $ respectively.  Comparison of these two figures indicates that accuracy of \eqref{Mod} through numerical calculation depends on the number of samples (training bit blocks).
According to Figure \ref{hy2} and \ref{hy}, the relation between $ \mathbb{P}_{b}^{[\textrm{apri}]}\left(b\right) $ and $ \mathbb{P}_{b}^{[\textrm{mod}]}\left(b\right) $ can be described by a shifted tangent hyperbolic function. Tangent hyperbolic function has also been applied as a tentative device for soft-decision making in iterative multi-user code-division multiple access decoders \cite{divsalar1998improved}.

We consider the following heuristic model as a probability transform function: 

\begin{equation}
\mathbb{P}_{b}^{\textrm{[mod]}}\left(+1\right)=\frac{\textrm{tanh}\left(\alpha\left(\mathbb{P}_{b}^{\textrm{[apri]}}\left(+1\right)-0.5\right)\right)+1}{2}
\label{tu}
\end{equation}
where $ \alpha $ is a tuning factor defining the slope of the function. The tuning factor need to be optimized through training sequence for each turbo-CS iteration number. A numerical non-linear regression method can be applied to find $ \alpha $ for each turbo-CS iteration with input training bit sequence. In this work, we use Levenberg-Marquardt nonlinear least squares algorithm as a method of curve fitting\cite{seber2012linear}. 

\begin{figure}[]
\psfrag{A}[][]{$ \mathcal{I}_{\textrm{CD}}^{[\textrm{apri}]} $, $ \mathcal{I}_{\textrm{SD}}^{[\textrm{ext}]} $ [bit]}
\psfrag{B}[][]{$ \mathcal{I}_{\textrm{SD}}^{[\textrm{apri}]} $, $ \mathcal{I}_{\textrm{CD}}^{[\textrm{ext}]} $ [bit]}
\hspace{-0.3cm}
\includegraphics[scale=1]{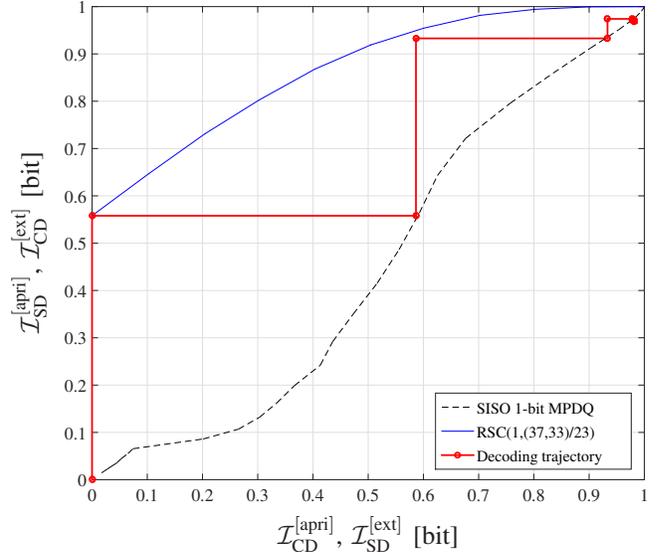}\caption{EXIT chart of turbo-CS receiver with modified output probabilities at SNR$=-1 $ [dB]. To obtain EXIT trajectory  $ \boldsymbol{\Lambda}_{\textrm{SD}}^{[\textrm{mod}]}$ is used instead of $ \boldsymbol{\Lambda}_{\textrm{SD}}^{[\textrm{ext}]}$. \label{EX2}}

\end{figure}

We obtain optimized $ \alpha $ for 3 turbo-iterations through $ T=50 $ blocks of training bit sequence with block size $ M=500 $. The decoding trajectory of turbo-CS decoder with refined SISO 1-bit MPDQ is depicted in Figure \ref{EX2}. With this probability refinement, the coding trajectory matches the EXIT functions.

\begin{figure}[t]
\psfrag{A}[][]{SNR [dB]}
\psfrag{B}[][]{RSNR [dB]}
\psfrag{data1}[][]{{\footnotesize Itr. 1}}
\psfrag{data2}[][]{{\footnotesize Itr. 2}}
\psfrag{data3}[][]{{\footnotesize Itr. 3}}
\psfrag{data4}[][]{{\footnotesize Itr. 4}}
\psfrag{data5}[][]{{\footnotesize Itr. 5}}
\psfrag{data6}[][]{{\footnotesize Itr. 6}}
\psfrag{TX}[][]{{\footnotesize SNR improv.}}
\psfrag{TR}[l][]{{\footnotesize RSNR improv.}}

\hspace*{-0.5cm}
\includegraphics[scale=1]{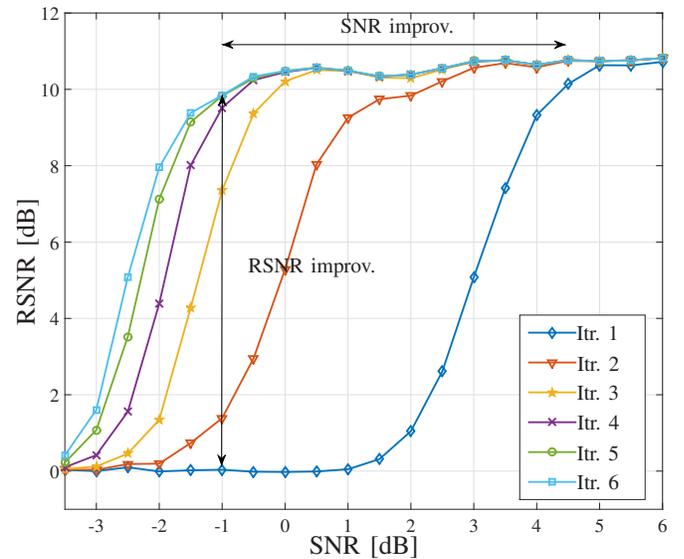}\caption{Performance of turbo-CS decoder for 6 iterations and for different SNR scenarios ($ N=1000 $, $ M=500 $, $ \rho=0.01 $)  \label{SNR}}
\end{figure}
\section{Performance of turbo-CS decoder}\label{per}
In this section, we briefly investigate the performance of turbo-CS decoder. We assume that signal $ \mathbf{x}\in \mathbb{R}^{N} $ is generated based on the distribution in (\ref{dis}) where $ N=1000 $ and $ \rho=0.01 $. The measuring matrix is $ \boldsymbol{\Phi}\in\mathbb{R}^{M\times N} $ with i.i.d zero-mean Gaussian entries with variance $ 1/M $ where $ M=500 $.  Note that $ \Phi $ is known by the turbo-CS decoder.

We apply a recursive systematic rate 1/3 convolutional encoder with polynomial coefficients $\mathbf{G}=\left(1,\frac{37}{23},\frac{33}{23}\right)_{8}$ as the channel encoder. The coded bits then pass through an AWGN channel. 

For turbo-CS tuning purpose, a sequence of $ T=50 $ predefined bit-blocks are input to the channel decoder. For each SNR the tuning factor $ \alpha $ in \eqref{tu} is derived for 6 turbo-CS iterations. 

Then turbo-CS receiver is applied to reconstruct the signal through 6 turbo-iterations. The maximum number of inner iterations is set to $ 100 $. We measure the performance of the signal reconstruction by received signal to noise ratio (RSNR) which is defined by
\begin{equation}
\textrm{RSNR}=\frac{\mathbb{E}\left(\left\Vert \mathbf{x}\right\Vert _{2}^{2}\right)}{\mathbb{E}\left(\left\Vert \hat{\mathbf{x}}-\mathbf{x}\right\Vert _{2}^{2}\right)}
\end{equation}
where $\left\Vert \mathbf{x}\right\Vert _{2}$ denotes the Euclidean norm of vector $ \mathbf{x} $.

Figure \ref{SNR}, illustrates the performance of turbo-CS receiver for 6 iterations over 4000 Monte Carlo realizations for different SNRs.
As it is depicted in the figure, turbo-CS decoder improves the quality of reconstructed signal by more than $ 10 $ dB at an SNR of $ -1 $ dB in comparison to non-iterative decoder (first iteration). Considering a fixed RSNR, we improved the SNR range by more than 5 dB when the turbo-CS decoder has converged (above RSNR of $ 10 $ dB). 

What we can also notice here is the typical turbo-decoder effects; Firstly, over iterations the performance improvement diminishes. Secondly, the slope of the curve is much steeper than our previous paper \cite{movahed2014iterative}, creating a turbo-cliff performance where the system either converges or it does not.

\section{Conclusion}\label{sec6}
In this paper, we introduced a serial concatenated encoding scheme with an iterative source-channel decoding method. The system model is designed for transmission of sparse signals over an AWGN channel which we refer to as a turbo-CS encoder. We proposed SISO MPDQ as a soft-in/soft-out source decoder that can be concatenated with an APP decoder to decode sparse signals in an iterative fashion, which we refer to as turbo-CS decoder.

We have shown, for the first time, the EXIT chart analysis of the turbo-CS decoder, we modified the soft-outputs of SISO MPDQ to be of the correct distribution required by the APP decoder. 

We considered the case where the sparsity level of the transmitted signal is stochastic and unknown to the turbo-CS decoder. The numerical experiment shows more than $ 10 $ dB improvement in the signal reconstruction performance (RSNR) can be achieved (at a channel SNR of $ -1 $ dB) over the non-iterative (first iteration) decoder. Alternatively, more than $ 5 $ dB of channel SNR can be improved for a fixed signal reconstruction performance of RSNR$ =10 $ dB.
\appendices
%
%
%




\ifCLASSOPTIONcaptionsoff
  \newpage
\fi

\section*{Acknowledgement}
The authors would like to thank Sundeep Rangan and Ulugbek S. Kamilov for providing us with their MPDQ software and answers to our questions on their method, which greatly helped us in determining this result. In addition, thanks to Ingmar Land for useful discussions and sage advice.
\balance
\bibliographystyle{IEEEtran}

\bibliography{IEEEabrv,reference}

\end{document}